# Temporal Cloak without Synchronization

*Zhixing Lin, Shuqian Sun, Wei Li, Ninghua Zhu and Ming Li*

*Abstract*—Considered to only exist in the fairy tales in the past, invisibility cloaks have been successively converted into reality no matter in the spatial domain or temporal domain. Inspired by the spatial cloaking, time gaps are utilized to hide temporal events. However, a sophisticated synchronization for cloaking is indispensable in these cloaking techniques, therefore leaving inconvenience for the realization of the temporal cloak. Here, by exploiting temporal Talbot effect, we propose a brand new scheme and concept to achieve a temporal cloak without any synchronization for cloaking process, under which the intensity-modulated event is directly turned into invisibility in intensity through temporal averaging effect induced by the Talbot effect. We successfully realize temporal cloak for periodic and pseudo-random signals respectively. We also find that the higher order temporal Talbot effect is beneficial to cloaking performance. Moreover, it is feasible to recover the data from the intensity-cloaked waveform. This method and its concept render a distinct perspective for temporal cloak, extend temporal cloak to the pulsed-wave and promote the development of confidential communication.

## I. Introduction

The idea of demonstrated temporal cloaks originally was derived from one of the schemes of spatial cloak[1-7] -- hiding the object in an artificial spatial "hole" [1, 2]. By manipulating the propagating direction of the probe light, the probe light converges or diverges in a specific fashion to bypass the object. Thus, no light interacts with the object and no object can be detected. This dramatic manipulation is mainly driven by the relevant advance in materials, including materials with negative refractive index [8, 9]. It is natural to extend the spatial cloaking concept to the temporal domain through space-time duality [10-13], and a new concept of frequency gap was recently reported to fulfill a full-field broadband invisibility [14]. In detail, the previously demonstrated temporal cloak imparts a quadratic phase profile in time domain, which is realized by the nonlinear process of four-wave mixing or phase modulation. Then, with proper dispersion, the probe continuous-wave with different wavelengths will be advanced or delayed and transform into pulsed wave, generating periodic time gaps in which the events can be concealed. However, these previous temporal cloaking techniques impose requirements for precise synchronization between the time gaps and events, which lead to inconvenience to switching between cloak-on and cloak-off. It is true that a precise synchronization up to attosecond [15] and a high-accuracy synchronization [16] have been obtained, but a cloak without the requisite of synchronization is still the pursuit of the practical and convenient cloak.

The temporal Talbot effect was previously utilized to implement optical clock recovery [17, 18]. Applying this mechanism to the temporal cloak, we propose a novel cloaking temporal cloak which is designed not to avoid interaction between the event and probe but to conceal the intensity event that has already carried by the probe. It is conceptually distinct from the previous cloaking techniques on the employed rationale, leading to elimination of the necessity to precisely synchronizing the temporal gaps and events when switching on or off the cloak, and to the recoverability of the concealed events.

## II. Principle

The theoretical foundation of the presented temporal cloak is the temporal Talbot effect. The Talbot effect was firstly discovered by Talbot in 1836 [19]. Later, the consistency of the mathematical expression between the diffraction in the spatial domain and the dispersion in the temporal domain, namely space-time duality, was found [20, 21]. Due to the space-time duality, many spatial concepts such as the thin lens and the Talbot effect, were transplanted to the time domain [22-24]. The Talbot effect in time domain mentions that the pulse train will be reproduced in different combinations of amplitudes and periods through the propagation of the dispersive medium with different dispersion coefficients. The dispersion necessitated in the temporal Talbot effect is usually provided by the linearly chirped fiber Bragg grating (LCFBG) or dispersion compensating fibre (DCF).

Unlike the continuous-wave temporal cloak, the proposed cloak is based on the magical characteristic of the temporal Talbot effect – the temporal averaging effect (TAE). The requirement of the TAE is that the probing light is comb-like in optical spectrum and each tooth is coherent in phase, which means that this cloaking method is only feasible for mode-locked pulses. Under the propagation through dispersive medium with a specific dispersion coefficient (integer Talbot dispersion coefficient), the spectral components of each pulse are retarded and advanced to the different timeslots with the different spectral deviations between the central wavelength and spectral teeth. The integer temporal Talbot dispersion coefficient $|\ddot{\beta}|L$ satisfies $sT^2=2\pi|\ddot{\beta}|L$, where $s$, $T$, $|\ddot{\beta}|$ and $L$ are a natural number, the repetition period, the second-order dispersion coefficient and the dispersive propagation length, respectively. In a certain temporal span, the time-shifting

This Manuscript received XXX X, XXXX; revised XXX X, XXXX. This work was supported by the National Natural Science Foundation of China under 61522509 and 61535012, M. L. was supported partly by the Thousand Young Talent Program. *Corresponding author: Ming Li, ml@semi.ac.cn.

Z. Lin, S. Sun, W. Li, N. H. Zhu and M. Li are with State Key Laboratory on Integrated Optoelectronics, Institute of Semiconductors, Chinese Academy of Sciences, P. O. Box 912, Beijing 100083, China and School of Electronic, Electrical and Communication Engineering, University of Chinese Academy of Sciences, Beijing 100049, China (email: zhxlin@semi.ac.cn; sqsun@semi.ac.cn; liwei05@semi.ac.cn; nhzhu@semi.ac.cn; ml@semi.ac.cn).

spectral component regenerates a new pulse which contains the spectral teeth of every neighboring-period pulse. Thus, the extension of the temporal Talbot effect to the amplitude-varied pulse train lays the foundation of the temporal cloak we proposed here.

The schematic diagram of the temporal cloak based on temporal averaging effect is presented in Fig.1. Fig.1 also shows its contrast – the visible status, where the dispersive medium is absent. When applying an event on the initially flat pulses, the event is transferred to the pulses. The dispersive medium ('cloaking device' shown in Fig.1) with a special dispersion coefficient $|\ddot{\beta}|L=sT^2/2\pi$, relocates the different spectral components (different color circles shown in Fig.1) of each amplitude-inconsistent timeslot to the neighboring timeslots. All the spectral components delayed to certain timeslots add coherently to form new pulses. Because the period of newly-formed pulse consists of ones of every neighboring-timeslot pulse's spectral components, the optical spectrums of the reconstructed pulses in different timeslots are approximately the same within a certain time span, which lead to the TAE consequently. In the process of the TAE, there exist many intermediate states (that is $s \cdot LT/n$, where LT is the integer Talbot distance, $s$ and $n$ are natural numbers) where the TAE has been achieved. However, the temporal cloak is invalidated by the mismatch between the repetition rate of output pulses and input pulses. Thus, in order to remain the repetition rate of the cloaked pulses unchanged, we select the integer Talbot effect as the exact cloaking case. By the TAE, a new conception of the temporal cloak construction is framed to flatten the intensity from the event-carrying probing pulsed light, differing from all the previous cloaks based "gaps". Due to the difference, specific features arise including the absence of synchronization for cloak and recoverability of data. Although the intensity is restored to the initial pattern, the dispersion process sacrifices the initiative constant profile. In other words, this cloaking method is an intensity-to-phase conversion process, which convert the intensity event into an undetectable to a regular photodetector. For further consideration, this new methodology may offer a temporal counterpart of the spatial cloaks which try to restore the contorted wave from object [3-7].

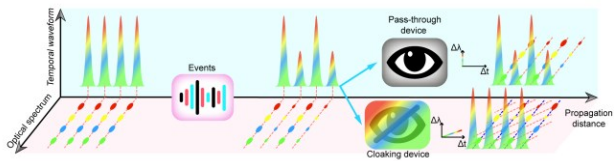

Fig. 1. Schematic diagram of the temporal cloak based on the temporal averaging effect. To be noticed, for the clarity of manifesting the mechanism of the TAE, the spectral components are drawn discretely, but actually the spectrum is a Fourier transform result of a pulse train, not a single pulse.

III. RESULTS AND DISCUSSIONS

Fig. 2(a) shows the experiment setup of our temporal cloak system. The actively mode-locked laser (AMLL) is employed as the optical source, whose repetition rate is controlled by the analog signal generator (ASG). Given that the original bandwidth of the AMLL is too wide, the second order dispersion coefficient of the dispersion fiber cannot be regarded as a constant. Hence, the optical bandpass filter (OBPF) with a bandwidth of 650 pm is employed to limit the optical spectrum in the system. Because of the discrete nature of the optical carrier, a tunable optical delay line (TODL) is necessary to align the carrier pulse train with the high-frequency modulating signal (events). The event is generated by an arbitrary waveform generator (AWG) or a parallel bit error ratio tester (PBERT). Before being applied on the intensity modulator (IM), it needs to be amplified by an electrical amplifier (EA). Then the modulated optical signal is amplified by an EDFA before entering the cloaking device -- a spool of dispersion compensation fibre (DCF) with a specific dispersion. It is worth noting that the pump power of EDFA has to be controlled properly to avoid nonlinear effects in the DCF. Thanks to the TAE in the DCF, the power of each pulse is equally distributed to adjacent pulses. The state of "cloak off" is simply acquired by removing the DCF. Eventually, the modulated pulse train is flatted in the time domain.

Here, the intrinsic differences between the alignment of the system we proposed and the synchronization of previous cloaks shall be particularly clarified: owing to the difference of cloaking concept, the alignment occurs in the process of modulation, whereas the synchronization occurs in the process of cloaking. This difference qualifies us to turn on or off the cloak freely without synchronizing the cloaking device. Yet, It cannot be specified as superiority over previous cloaks, but a unique feature.

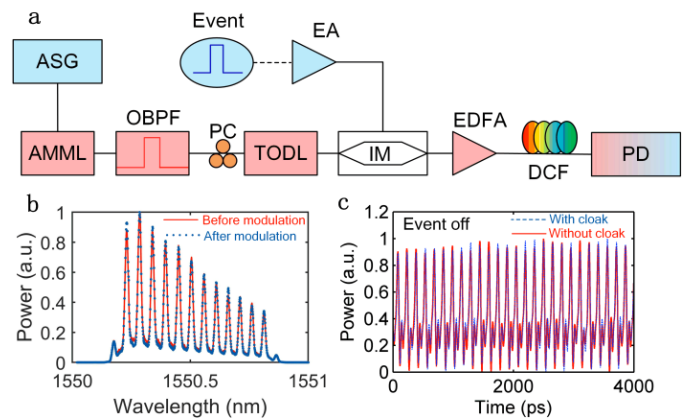

Fig. 2. Experimental configuration of the proposed temporal cloaking system. ASG, analog signal generator; AMLL, actively mode-locked laser; OBPF, optical bandpass filter; PC, polarization controller; TODL, tunable optical delay line; IM, intensity modulator; EA, electrical amplifier; EDFA, erbium-doped fibre amplifier; DCF, dispersion compensation fibre; PD, photodetector. The temporal events are generated by an arbitrary waveform generator (AWG) or a parallel bit error ratio tester (PBERT). (b) Spectra before (solid red) and after (dot blue) the modulation of a 3.305-gigahertz squarewave at a resolution of 0.01 nm. (c) Measured temporal waveforms with (solid blue) and without the temporal cloak (dashed red) when the event is off. Few discrepancies can be seen.

Generally, intensity modulation will generate evident sidebands beside a carrier. And these sidebands are undesired in temporal cloaking systems. Fortunately, with a designed extinction ratio the pronounced sidebands are not observed. As shown in Fig. 2(b), the spectra with (blue dot) and without (red solid) modulation are almost the same where the corresponding extinction ratio of events is ~50% (a higher extinction ratio leads to visible sidebands in experiment). While low sidebands exist in MATLAB simulation, in experiment the perturbation and imperfections of the probing light make these low sidebands more indistinguishable, leading to the conformity of the spectrums before and after a suitable modulation. For a temporal cloaking system, the invariability of the optical spectrum is desired. Our system experimentally meets this requirement through the trade-off between the extinction ratio and the undetectable optical spectral distinction. Thanks to these features, a path to temporal cloak is paved. Specially, no synchronization for cloaking process is required in our temporal cloaking system.

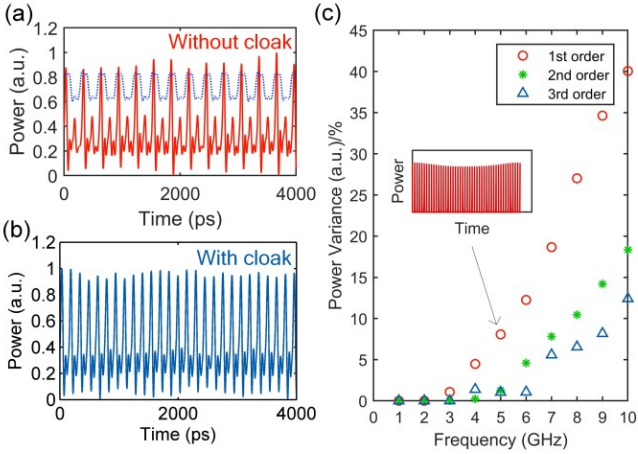

Fig. 3. Cloaking performance of periodic events. (a) Measured temporal waveforms of the modulated pulse sequence at a repetition rate of 6.61 gigahertz and the periodic voltaic signal (dashed blue, not relative to power values shown on the y-axis) at a data rate of 6.61 Gbps. (b) The measured temporal waveform after cloaking. (c) Simulation of the power variances of the different order averaging effects in the different ratios between the event data rate and the pulse repetition rate. Inset: the cloaked waveform when the ratio is 1/5. The power variance is defined as the difference between the maximum and the minimum values of the pulse peak values and is normalized to the maximum of pulses for the sake of comparison.

The experimental and simulation results of periodic events are shown in Fig.3. As can be seen, it is hard to distinguish the cloaked waveform [Fig. 3(b)] from the "raw" waveform in the cloaking experiment for the periodic event. The dispersion value used here is ~-2850 ps/nm, exactly matching the repetition rate – 6.61 gigahertz which satisfies the equation of the first order integer Talbot effect. The fluctuation emerging on the cloaked waveform might be caused by the slope of dispersion coefficient or the unsteadiness and optical spectral imperfection of the AMLL. Due to the immutable dispersion value provided by the DCF, we perform a simulation to figure out the influence to the cloaking perform made by the data rate and the order of averaging effect. In this simulation, the data rate of the periodic event is set at 1 Gbps, e.g. 1010……, and the extinction ratio is 50%, while the repetition rate of pulses is ranging from 1 to 10 gigahertz. The ratio varying from 1 to 1/10 enables it to investigate the cloaking tolerance for the ratio lower than 1. The influence by virtue of the order of the Talbot effect is presented in Fig.3(c). As the ratio goes down with the repetition rate of pulses arising, the increasing variance indicates that the cloaking perform is aggravating. To be noticed, when the ratio for 1st, 2nd and 3rd descends to 1/5, 1/7 and 1/9, respectively, the envelope of the pulses cannot be removed by this cloaking technique, namely, the temporal cloak become invalid at their cutoff ratio. Hence, for our cloak, the ratio of the data rate of event to the repetition rate of pulse train should not be lower than the cutoff ratios for the different order TAEs. Evidently, the higher order of averaging effect reduces the amplitude variance, which makes the cloaking perform better. It can be ascribed to the multiple distributing temporal range nΔt because of the multiple dispersion value $n|\ddot{\Phi}|$, where n is the order of the TAE. However, the higher order TAE puts a more precise requirement for the correspondence between the repetition rate of pulses and the dispersion value, since the deviation of distributing temporal distance Δt is amplified n times at the same time.

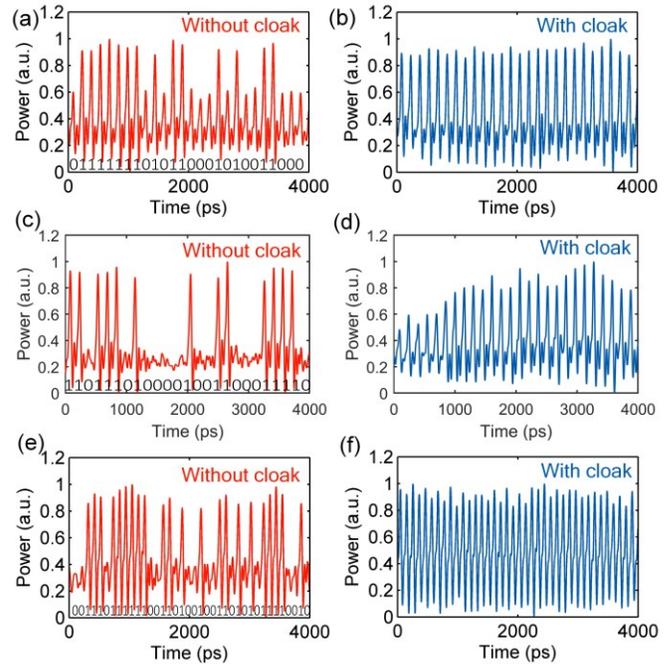

Fig. 4. Cloaking of pseudorandom signals. (a)/(c) Measured pulse sequences modulated by the pseudorandom signals with a low and high modulation index, respectively. Its repetition rate is 6.61 gigahertz satisfying the condition for the first order Talbot effect. (b)/(d) Cloaked signals of (a)/(c) after going through the cloaking element. (e) Measured pulse sequence modulated by the pseudorandom signal with a higher modulation index. Its repetition rate is 9.61 gigahertz satisfying the condition of the second order Talbot effect. (f) Cloaked signal of (e) after undergoing the cloaking element.

To determine the cloaking capability for data, the pseudo-random signal is exploited as the modulating signal, and relevant results are given in Fig.4. In the case where the dispersion coefficient is ~-2850 ps/nm, the repetition rates of Fig.4(a)/(b), Fig.4(c)/(d), and Fig.4(e)/(f) are 6.61 gigahertz, 6.61 gigahertz, and 9.61 gigahertz, satisfying the first order Talbot effect, the first order Talbot effect, and the second order Talbot effect, respectively. From the experimental results shown in Fig.4(a)/(b) and (e)/(f), our temporal cloak makes the pseudo-random signal absent from the pulse train. As mention in previous paragraph, the higher order TAE results in the better performance. Fig. 4(c)/(d) are presented to partly confirm the conclusion. With a higher extinction ratio, the first order TAE cannot achieve an adequate cloaking waveform like Fig. 4(a) and (b) do. However, the second order TAE [shown in Fig. 4(e)/(f)] qualifies the system for the cloak of high-contrast signal. The better performance owing to a wider averaging range is achieved according to the second order TAE, which confirms the above statement that the higher order leads to the better cloaking performance. As for the perturbation of the pulses, it is potentially caused by the reasons described above.

Whether for periodic or non-periodic digital signals, our temporal cloaking technique presents a desirable cloaking performance. But the price we pay for the intensity invisibility is the change of phase. It is also possible for cloaking analog signals, as long as the modulation depth, the frequency of the analog signal, the repetition rate and the bandwidth of the pulse train are appropriately selected. A cascaded TAE may improve the performance of cloak because it can distribute the spectral components to more adjacent pulses. The only factors that affect the performance are whether the optical spectrum of AMLL's output is a perfect comb function and the performance of dispersion medium. Hence, the extinction ratio of optical spectrum has to be high, otherwise pulses will diffuse throughout the temporal domain. Replacing the DCF with an LCFBG, the cloaking system will be more compact and the TAE will perform better thanks to its uniform dispersive response of the LCFBG. In this way, the optical filter is also unnecessary since the dispersion coefficient is wavelength-independent. With the accessibility of arbitrary dispersion values, the repetition frequency of the pulse train can be arbitrarily set at the frequency that an actively mode-locked laser supports, making our temporal cloaking method meet the cloaking requirements for the high-speed data. Apart from utilizing the LCFBG, the application of optical wave shaper might have positive effect on temporal cloak through flatting the optical spectrum of pulses. Besides, our temporal cloak has the ability of recovering the information from the cloaked waveform through the opposite dispersion. If it is undesired to recover the data easily through measuring the repetition rate of pulses, we can apply a designed seeming-arbitrary phase profile to the pulse train, which might also help to restore the imported phase change. From this perspective, we would say that the event is concealed under the cloak rather than disappears.

## IV. Conclusion

Using the TAE, we present a new scheme for the temporal cloak to conceal the event carried by the probing light with the absence of synchronization, and experimentally demonstrate the data-recoverable temporal cloak for the periodic and pseudorandom signal in different repetition rate. Additionally, through simulation and experiment, we confirmed that the performance is able to be improved by increasing the order of the integer Talbot effect. With such a simple system, we can obtain similar results—making event undetectable to intensity detection—compared with the previous cloaks. The proposed technique holds the potentiality of benefiting for practical optical secure communication and proposing a new route for the temporal cloak.